\pdfoutput=1
\documentclass[namedreferences]{solarphysics}

\usepackage[optionalrh,solaromanenum]{spr-sola-addons} 
\usepackage{hyperref}
\hypersetup{colorlinks=true,
linkcolor=black,
citecolor=black,
urlcolor=blue,
breaklinks=true}
\usepackage{graphicx}                    
\usepackage{color}                       
\usepackage{breakurl}                         

\newcommand{\eg}{\textit{e.g.}}

\newcommand{\etal}{\textit{et~al.}}
\newcommand{\etc}{\textit{etc.}}

\begin{document}

\begin{article}

\begin{opening}

\title{The greatest GOES soft X-ray flares: Saturation and recalibration over two Hale cycles}

%
 \author[addressref={1,2},corref,email={hugh.hudson@glasgow.ac.uk}]{\inits{}\fnm{Hugh}~\lnm{Hudson}\orcid{0000-0001-5685-1283}}
 \author[addressref={2,3},corref,email={}]{\inits{}\fnm{Ed}~\lnm{Cliver}\orcid{}}
\author[addressref={4},corref,email={}]{\inits{}\fnm{Stephen}~\lnm{White}\orcid{}}
\author[addressref={5,6},corref,email={}]{\inits{}\fnm{Janet}~\lnm{Machol}\orcid{}}
\author[addressref={5,6,7},corref,email={}]{\inits{}\fnm{Courtney}~\lnm{Peck}\orcid{}}
\author[addressref={8},corref,email={}]{\inits{}\fnm{Kim}~\lnm{Tolbert}\orcid{}}
\author[addressref={5,9},corref,email={}]{\inits{}\fnm{Rodney}~\lnm{Viereck}\orcid{}}
\author[addressref={10},corref,email={}]{\inits{}\fnm{Dominic}~\lnm{Zarro}\orcid{}}

%
\runningauthor{Hudson \textit{et al.}}
\runningtitle{Greatest GOES events}

\address[id={1}]{SSL, UC Berkeley, CA USA}
\address[id={2}]{School of Physics and Astronomy, University of Glasgow, UK}
\address[id={3}]{National Solar Observatory, 3665 Discovery Drive, Boulder CO 80303 USA}
\address[id={4}]{Air Force Research Laboratory, Albuquerque NM USA}
\address[id={5}]{Cooperative Institute for Research in Environmental Sciences (CIRES), CU Boulder}
\address[id={6}]{National Centers for Environmental Information (NCEI), NOAA}
\address[id={7}]{Laboratory for Atmospheric and Space Physics (LASP), 3665 Discovery Drive, Boulder CO 80303 USA}
\address[id={8}]{American University and NASA/GSFC}
\address[id={9}]{Space Weather Prediction Center (SWPC), NOAA}
\address[id={10}]{ADNET Systems, Inc.}

\begin{abstract}
The solar soft X-ray observations from the GOES satellites now span two full Hale cycles and provide one of the best quantitative records of solar activity, with nearly continuous flare records since 1975. 
We present a uniform analysis of the entire time series for 1975 to 2022 at NOAA class C1 level or above, to characterize the occurrence distribution function (ODF) of the flares observed in the 1-8~\AA\ spectral band.
The analysis includes estimations of the peak fluxes of the 12 flares that saturated the 1-8~\AA\  time series.
In contrast to the standard NOAA classifications, these new estimates use the full time resolution of the sampling and have a pre-flare background level subtracted.
Our new estimates include NOAA's recently established correction factor (1.43) to adjust the GOES-1 through GOES-15 data covering 1975-2016.
For each of the 12 saturated events we have made new estimates of peak fluxes based on fits to the rise and fall of the flare time profile, and have validated our extrapolation schemes by comparing with artificially truncated but unsaturated X10-class events.
In this new estimation, SOL2003-11-04 (the most energetic unambiguously observed event) has a peak flux of  4.32 $\times10^{-3}$\,W/m$^2$.
This corresponds to X43 on the new scale, or X30 on the old scale.
We provide a list in the Appendix for peak fluxes of all 38 events above $10^{-3}$~W/m$^2$, the GOES X10 level, including the 12 saturated events.
The full list now gives us a first complete sample from which we obtain an occurrence distribution function (ODF) for peak energy flux  $S$, often represented as a power law $dF/dE \propto E^{-\alpha}$, for which we find $\alpha = 1.973 \pm 0.014$ in the range M1 to X3.
The power-law description fails at the high end, requiring a downward break in the ODF above the X10 level.
We give a tapered powerlaw description of the resulting CCDF (complementary cumulative distribution function) and extrapolate it into the domain of ``superflares,'' \textit{i.e.} flares with bolometric energies $>10^{33}$~erg.  
Extrapolation of this fit provides estimates of 100-yr and 1000-yr GOES peak fluxes that agree reasonably well with other such estimates using different data sets and methodology, although there is some tension between our 10,000-yr (the Holocene time-scale) estimate and the GOES class  obtained for the out-sized 774~AD solar proton event as inferred from cosmogenic nuclide records. 
\end{abstract}

%

\end{opening}

%
 \section{Introduction}\label{sec:intro} 

The A,B,C,M,X soft X-ray (SXR; 1-8 \AA) flare classification system based on observations by the  Geostationary Operational Environmental Satellite (GOES) network has become the standard measure of solar flare strength.  
These data come from the X-ray Sensor (XRS) instruments on board these satellites, and in this paper we routinely refer to GOES/XRS data as simply GOES data.
The classification scheme is defined as follows: GOES classes A1-9 through X1-9 correspond to flare peak 1-8 \AA\ fluxes of $(1-9) \times 10^{n}$~Wm$^{-2}$ where $n = [-8, -7, -6, -5, -4]$, for classes A, B, C, M, and X, respectively. 
Occasionally, flares are observed with peak fluxes $\geq 10^{-3}$~Wm$^{-2}$.  
Rather than being assigned a separate letter designation, such flares are referred to as $\geq$X10 events; it is these few events (often ``saturated'' ones)\footnote{``Saturation'' is a common misnomer here; the sensors have a linear response and the clipping that is observed results from the range limit of the data numbers. See Section~\ref{sec:corrections} for details.} that we focus on here.
Traditionally, the peak-flux class assigned a given event comes from the largest one-minute average of the flare's time profile without correction for background.

Prior to the GOES spacecraft (the first of the series launched in 1975), similar flare soft X-ray measurements were initiated on a series of SOLRAD and other early spacecraft such as the two SMS satellites, beginning as early as 1963 \citep{1991SoPh..133..371K}.
\cite{2011SoPh..272..319N} describes the earlier GOES data and the intercalibration issues of the GOES spacecraft up through GOES-14.
We do not include any of the related data from missions prior to GOES-1 in our present treatment for various reasons (weak metadata, inaccessibility, technology differences, database incompleteness, \textit{etc.}).
During the long history of the GOES/XRS instruments,  the two major technical changes were the switch from spin-stabilized to 3-axis spacecraft with GOES-8 \citep{1996SPIE.2812..344H}, and the detector change from ion chambers to Si photodiodes beginning with GOES-16 \citep{2009SPIE.7438E..02C,2020bookMACHOL2020233M}.

In the switch from spin-stabilized satellites to 3-axis-stabilized satellites, the XRS measurements were found to disagree, with GOES-8 through GOES-15 measuring higher irradiances than the preceding instruments. 
To maintain consistency with previous satellites, the NOAA Space Wea\-ther Prediction Center (SWPC) used ``SWPC scaling factors'' to adjust the GOES-8 through GOES-15 XRS measurements to match the previous measurements
\citep[][provide a full description of this]{2022GOESrept.....M}.
These corrections were applied to all GOES-8 through GOES-15 XRS measurements and are commonly known as the ``SWPC scaling factors". 
Based on careful pre-launch calibrations of the GOES-16 and GOES-17 XRS instruments, NOAA determined that the GOES-8 through GOES-15 instruments had the correct calibrations, and upward scaling should have instead been applied to the earlier GOES-1 through GOES-7 satellites\footnote{The GOES-1 and GOES-2 were calibrated using a solar blackbody spectrum, rather than the flat spectrum adopted later; the database will be corrected in the future. This will slightly affect the results in this paper for SOL1978-07-11 in the Appendix.}.
For GOES-8 through GOES-15, NOAA's operational GOES irradiances and flare classifications have used XRS data with the SWPC scaling factor applied, resulting in flare magnitudes that are approximately 40\% lower than the modern correct values. 
To properly calibrate GOES-1 through GOES-15 irradiances \textit{as reported prior to these corrections}, all irradiances for operational GOES-1 through GOES-15 should thus be multiplied by a scaling factor of 1.43 and the flare classifications should correspondingly be adjusted to match the irradiances.
Reprocessing of a new science-quality version of the GOES-1 through GOES-15 data with the calibration corrected is underway at the NOAA National Centers for Environmental Information (NCEI); it is complete for GOES-13 through GOES-15 at the time of writing and will be fully implemented in 2023 for the earlier satellites, making the eventual database consistent with the current GOES-R series. 
This paper only deals with XRS Channel--B data (1--8~\AA), and concludes with a table of events at $\geq$X10  in this corrected modern scale.

Since the beginning of GOES SXR observations in 1975, there had been 22~events listed with a classification $\geq$X10.0, and (depending upon epoch) such events may have driven the sensors' readings into saturation. 
The greatest such event, SOL2003-11-04, was assigned an estimated peak value of X28,  though it saturated at the X18.4 level.\footnote{We have found no documentation as to how this flare magnitude was determined. 
For most of the other $\geq$X10 flares, the reported GOES peak flux corresponds to a value at or slightly above the saturation level.}
The earlier SOL2003--10--28 event saturated only slightly at the X18.4 level in GOES-10 (but did not saturate at X17.2 in GOES-12) and probably served as a template for the class of the bigger event \citep{2004AAS...204.4713K}.

Given the susceptibility of our electronic infrastructure to space weather \citep[e.g.,][]{2013SpWea..11..138C,2021SpWea..1902593H}, the magnitudes of the most powerful GOES SXR flares have great practical interest.  
While the 1-8~\AA\  SXR band contains only $\leq 2$\% of the flare radiative or bolometric energy \citep{2006JGRA..11110S14W}, the GOES SXR class correlates (but not linearly) with estimates of the total flare radiative energy \citep{2011A&A...530A..84K}.  
The GOES class also can be used as an indicator of the potential impact of associated geomagnetic storms and solar energetic proton (SEP) events -- ``potential,'' because the effects of these phenomena depend on considerations other than flare magnitude, \textit{e.g.,} the occurrence of an associated  coronal mass ejection (CME) and location of the flare on the Sun for both storms and (SEP) events and the direction of the magnetic field in the CME for magnetic storms.  
Moreover, the fact that the saturated GOES events constitute the high-energy tail of the occurrence distribution function (ODF) of these bursts makes them critical for extrapolation of the ODF to larger peak fluxes. 
The threat of solar ``superflares'' at this extreme end of the distribution has rightly attracted great interest  (\eg, Cliver \etal\ 2022), and improved values will inform the 100--year extreme-event Benchmarking exercise carried out by the US Space Weather Operations, Research, and Mitigation Subcommittee (SWORM 2018, IDA 2019).
In this study we incorporate our revised fluxes for the 12 saturated GOES SXR flares (from 1978 to the present) to assess the ODF across its entire range.

In this paper we analyze GOES/XRS only on the basis of the existing user database as extracted via the the common data environment SolarSoft \citep{1998SoPh..182..497F}  (Section~\ref{sec:db}), and only for the 1--8~\AA\ channel (XRS--B, low photon energy, long wavelength).
In Section~\ref{sec:corrections}, we describe and implement our techniques for estimating peak 1--8 \AA\ fluxes for the saturated flares and in Section~\ref{sec:ofd} we incorporate these revised values in the cumulative frequency distribution of peak fluxes for the entire flare GOES database since 1975.  
This represents the first complete sample of this standard proxy to total flare energy, with significance not only for space weather but also for analogous stellar events.
In the Appendix we provide adjusted classes for the most energetic events, accurate to an estimated 20\%.

\section{Database}\label{sec:db}

\subsection{History and recalibration}\label{sec:recal}

Over the decades of GOES/XRS data, the instrumentation had evolved only slightly up until the GOES-R series  \citep[the current GOES-16, GOES-17, and GOES-18 at present, and GOES--19 in the future;][]{2020bookMACHOL2020233M}.
The GOES sensors, both current and earlier, essentially integrate all of the incident or ambient ionizing radiation.
This includes not only solar X--rays but any other source of ionization in the satellite environment.
The GOES-R \citep{2009SPIE.7438E..02C} solid-state photodiodes have closely similar X-ray responses but substantially different background properties, including greater sensitivity to ambient fluxes of energetic electrons.
\cite{1985SoPh...95..323T}, \cite{1994SoPh..154..275G}, and \cite{2005SoPh..227..231W} definitively discuss the calibration issues of GOES data up through GOES-12, including a detailed treatment of the effects of spectral model uncertainties such as elemental abundances in thermal emission spectra.
GOES-13 through GOES-17 data come from \url{https://www.ngdc.noaa.gov/stp/satellite/goes-r.html}, which includes several different background levels.
In this paper we uniformly used our own algorithm for individual flare background estimates.
These earlier papers do not discuss the calibration effect of the change from spin--stabilization to three--axis stabilization between GOES-7 and GOES-8, an example of the systematic issues that remain open at present (see Neupert, 2001\nocite{2011SoPh..272..319N}, who remarks that the spectral response functions for the spin--stabilized instruments do, however, incorporate the geometry of the scanning motion).

\begin{figure}[htbp]
\centering
   \includegraphics[width=\textwidth]{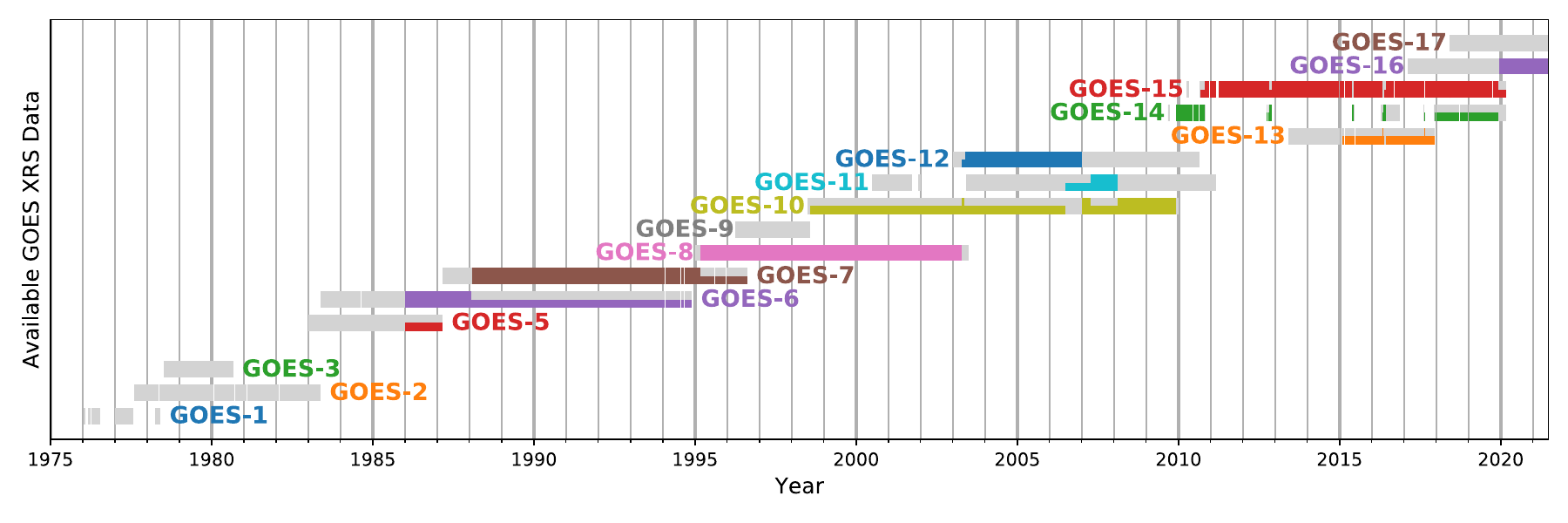}
            \caption{\textit{History of GOES/XRS data availability covering 1975 to 2022, showing primary (colored) and secondary (gray-shaded) sources where this information is available \citep[from][]{2022GOESrept.....M}.} 
      }
\label{fig:history}
\end{figure}

Our approach here is to analyze the irradiance data that NOAA make available at the time of writing, rather than to work with the original raw data (``counts,'' or raw digital signal levels) and attempt a comprehensive re-calibra\-tion. 
Indeed, much of this primary data has not been preserved.
The database work begins with event identification via the NOAA text files created incrementally in near real time and available on the Web\footnote{\url{https://www.ngdc.noaa.gov/stp/space-weather/solar-data/solar-features/solar-flares/x-rays/goes/xrs/}}.
We then used the SolarSoft {\sc ogoes} software to access the recalibrated data for each event, noting that the GOES-1 through GOES-15 data now include further cleaning, reflagging, and correction for temperature and digization issues.
The geosynchronous orbits of the GOES satellites result in a very high duty cycle, which exceeds 90\% even early in the program, and recently is near 100\%, with overlapping data streams from the three new-series GOES-R spacecraft at present.
Figure~\ref{fig:history} shows the time ranges of the various GOES spacecraft.

\begin{figure}[htbp]
\centering
   \includegraphics[width=0.9\textwidth]{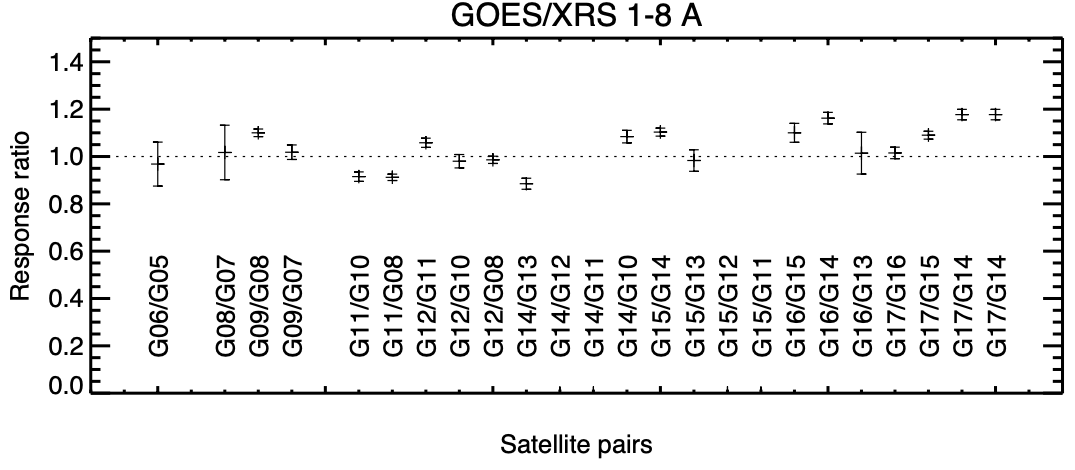}
          \caption{\textit{The average ratios of different pairs of GOES/XRS 1--8~\AA\ peak fluxes derived from simultaneous overlapping observations, using only events at the C1 level and above. 
Here ``G06'' stands for ``GOES-6'', \etc~The uncertainty flags show the standard deviations of the mean ratio.} 
      }
\label{fig:rats_plot}
\end{figure}

We have carried out a systematic intercomparison of overlapping GOES data streams, as summarized in Figure~\ref{fig:rats_plot}.
Not all such possible pairs existed, especially at earlier times.
This exercise essentially intercompares the independent absolute calibrations, as carried out prior to launch, of each of the spacecraft.
The GOES-16/GOES-15 ratio compares the newer and older detector technology, and the ``G16/G15'' point on the plot shows a ratio $1.10 \pm 0.04$ based on peak fluxes in the 1-8~\AA\ channel for 243 $\geq$C-class flares from occurring in 2017 to mid-2019.

Based on these comparisons, we conclude that the GOES radiometry has remained roughly stable over the decades, and the data have remained generally intercomparable from instrument to instrument to within about 20\%.
At this level of absolute calibration, we believe that future refinements and adjustments of the database will not significantly affect the conclusions we draw regarding the ODF.

The NOAA reports, in the text files noted above, defined the original ABCMX classifications.
These reports have many blemishes and ambiguities, and some systematic issues\footnote{\cite{2018SpWea..16..660S} note, for example, that there was no consistent method for determining flare onset time or end time, prior to 1997.}.
The NOAA text files (Flare Reports) also do not remove the scaling factor.
We note that NOAA plans to make corrected scalings available in these reports, extending back to GOES-1, in 2023.
Notably the classifications systematically overestimate the irradiances of the flares themselves, 
because they do not correct for the solar background signal levels. 
This makes them unsuitable for physical analysis, as noted by \cite{1990ApJ...356..733B} and others.
Our analysis of the data includes event-by-event background corrections in order to avoid this issue.
This work thus parallels that of \cite{2012ApJ...754..112A}, though with a different algorithm and with an additional decade of data accumulation \citep[see also][]{2022Physi...5...11S}.
While doing this background correction, we have also tried to eliminate as many as possible of the errors in the database as well, making checks to reject outliers in various parameters. 

For each event in the NOAA listing, we retrieved the full-resolution data from the source identified by the {\sc ogoes} tool in SolarSoft, using a fixed time window starting 1800~s prior to the NOAA listing of peak time for each flare, and continuing for 600~s afterwards.
At the time of writing, this access tool returns data that have the scaling factor removed for GOES-13 through GOES-15, but the earlier listings require the scaling factor to be removed manually until the NCEI repository is updated.

We estimate the peak time and peak flux from the fine sampling of the data (at 3-s, 2-s, and 1-s intervals depending on spacecraft), after finding background intervals algorithmically.
The algorithm locates a minimum level, usually in the half hour prior to the flare peak, for each flare.
Note that our use of the full time resolution available makes our numbers slightly inconsistent with earlier descriptions based on 1-minute averages.
We do this to improve (slightly) the physical significance of the results, noting that the background corrections have a more important systematic effect.
This comparison required screening against outliers, bringing the total number of C-class and greater down from 56,356 to 48,131-- a reduction of about 15\% -- covering the events from SOL1975-09-01T15:12 to SOL2022-06-15T07:25.
This cleaning-up of the database included visual inspection of all of the X-class events (as originally labeled), rejecting flawed data from which reasonable assessment of peak or background level could not be made, and the the case-by-case background correction.
We note also that the scale corrections and the variability of the background levels confuse the identification of C-class flares (see  \nocite{2021ARA&A..59..445H}Fig.~12 of Hudson, 2021, and its accompanying explanation). 

\subsection{Correction of flare classifications for saturated events}\label{sec:corrections}

The database contains 22 events at the $\geq$X10 level, of which 12 had saturated peak flux levels (clipped peaks) in the 1-minute high-resolution 1-8~\AA\ Channel~B data (note the 0.5-4~\AA\ Channel~A also saturated in a different and overlapping set of events, but we are not considering these data in our analysis).
The ``saturation'' we see in these 12~events (all prior to GOES-13) comes from the readout electronics, specifically in the range limit of digitization of the basic analog measurements.
We have no reason to suspect non-linearity of the analog signal near the digital thresholds, which vary from satellite to satellite.
The actual conversion from detector currents to physical units requires knowledge of the spectral distribution, which differs from flare to flare; this leads to some unavoidable systematic error \citep[e.g.,][]{1972SoPh...22..492W}.
The XRS readouts on the GOES-R (GOES-16 through GOES-18) satellites have a broader dynamic range than those on the earlier satellites, and so will not experience saturation events as easily. 
For instance, the XRS-B channel on GOES-16 will only saturate for an $\geq$X500 flare (at which point there are other things to worry about!).

We have used two techniques to make plausible extrapolations to determine best values for the peak fluxes of these events, testing our methods against artificially truncated X10-class flares for which we have complete time series; the database for this verification activity contains 10 such events (Table~\ref{tab:test}).
The GOES data sampling for these events is  is nominally 3-s (with sporadic intermittent occurrences of 2 and 4 s and occasional short gaps of ~10 s between data points), and we have used this resolution to estimate the peak values using two methods\footnote{The 12~saturated events all had nominal 3-s sampling, but with GOES-13 through GOES-15 the cadence increased to 2~s, and now with GOES-R it is 1~s.}.
The \textit{linear} method, arrived at by visual trial and error, crudely extrapolates the pre- and post-saturation time series (over 30-s intervals) at their full time resolution via linear extensions, determining the peak flux by the value at the intersections of these two lines.
The alternative \textit{functional} method makes use of a specific parametrized function,  fitted by eye in each case.
We emphasize that both methods are necessarily subjective, as with earlier similar efforts \citep[e.g.,][]{2004AAS...204.4713K}; here the common assumption between our two methods is that they rely upon the slopes of the light curve just before and just after the interval of saturation.

\begin{table}[h]
\caption{Test Events}
\label{tab:test}
\begin{tabular}{l l l l ll}
\hline
Event peak time & Class$^*$ & Cut$^*$ & Linear & Functional  & Satellite \\
\hline
SOL1982-06-06T16:37 & X10.1   &  X5  &  X9.3  & X8.4 &  GOES-2 \\
SOL1982-12-15T02:02 &  X12.9  &  X5  &  X10.0   & X9.0 &    GOES-2 \\
SOL1982-12-17T18:57 &  X10.1  &   X5  & X8.4  & X8.2 &  GOES-2  \\
SOL1984-05-20T22:36 &  X10.1  &   X5  & X9.2  &  X9.2 &   GOES-5 \\
SOL1991-01-25T06:34 &  X10.8  &   X5  & X10.0   & X10.7 &   GOES-6 \\
SOL1991-06-09T01:43 &  X10.5  &   X5  & X9.8   & X8.0  &   GOES-6 \\
SOL2001-04-15T13:49 &  X15.9  &   X8  & X16.0 & X16.4 &  GOES-10 \\
SOL2003-10-28T11:10 &  X18.6$^a$  &   X9   & X16.2 &X16.0 & GOES-10  \\
SOL2003-10-29T20:49 &  X10.9$^b$  &   X5   & X10.9 &X10.7 &  GOES-10 \\
SOL2005-09-07T17:40 &  X18.1$^c$  &   X9   & X19.7 &X16.9 &  GOES-12 \\
 \hline
 \end{tabular}
 
$^*$Determined from 3-s data \\
$^a$Saturated at $\sim$X18.4 for 1.4 min\\
$^b$Saturated at $\sim$X10.9 for $\sim$40 s\\
$^c$Blended time profile above saturation level?
\end{table}

\subsubsection{Linear fits}

This method relies on bridging the gap between the last and first good measurements at the ends of the saturation interval, by upward linear extrapolation of the last part of the flare rise curve prior to saturation, and the first part of the flare rise curve after saturation extrapolated to their point of intersection.

To see if this method had validity, we first tried it by artificially truncating the time profiles of a sample of large (near X10 or greater) GOES events (Table~\ref{tab:test}) at a level of about half the peak flux and then manually extrapolating from the remaining profiles on either side of the ``saturation'' gap to a peak flux.  
When this simple method showed promise -- the estimated peaks were reasonably close to the actual values -- we followed a more quantitative and reproducible approach:

\begin{enumerate}
\item fit a straight line through the last 30 seconds (10 data points) prior to saturation of the rise portion of an event;
\item do the same for the first 30 seconds data points after saturation on the decay;
\item extrapolate both lines upward until they intersect; and thence
\item obtain a one-minute average for the 20 highest data points contiguous to the point of intersection.
\end{enumerate}

\begin{figure}[htbp]
\centering
    \includegraphics[width=\textwidth]{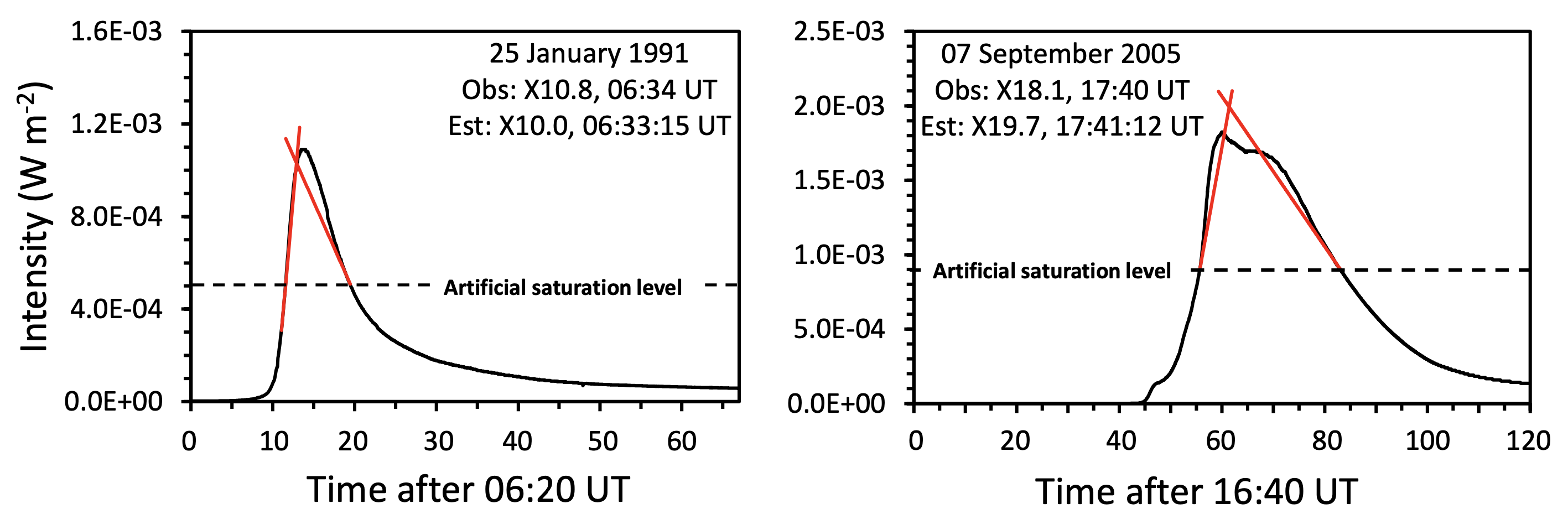}
         \caption{\textit{Left, estimation of flare GOES class (X9.9) for an artificially saturated X10.8 flare SOL1991-01-25. 
         Right, estimation of flare GOES class (X19.7) for an artificially saturated X18.0  event SOL2005-09-07. 
         Here the flare complexity compromises the peak estimation. }
      }
\label{fig:f1}
\end{figure}

This procedure is illustrated for SOL1991-01-25 in Figure~\ref{fig:f1} (left) and the results for all 10 events are given in the ``linear'' column of Table~\ref{tab:test}. 
In general the slope on the rise time is too steep (because it does not allow for the decrease of the rise rate as one approaches the peak) and too shallow on the decay which is typically more exponential.  
These two errors can be somewhat offsetting, resulting in an earlier and more distinct peak which tends to precede that of the actual flare.   
In the Table it can be seen that the estimated peaks range from 78\% to 109\% of the actual peaks with a median value of 92\%.  
Figure~\ref{fig:f1} (right) shows that the method can fail if the time profile exhibits complexity above the saturation level;  SOL2005-09-07 was the only event with a ratio of the estimated-to-actual peak flux significantly greater than unity via this method.  

The linear-fit results for the 12  flares that actually saturated the GOES detector are given in Table~\ref{tab:meth}. 
For five of these events the estimated peak value is a factor of 1.5-2 larger than the nominal (archived) value. 
For SOL2003-11-04 we obtained a revised classification of X30.5. 
This approach (the linear fit) is similar to that of \cite{2004AAS...204.4713K} who obtained a peak of X30.6.
In an abstract of an AAS talk, \cite{2004AAS...204.4713K} mention comparing the time profile of this event with other flares from the ``Halloween'' episode region (NOAA 10486; Gopalswamy \etal, 2005) that produced SOL2003-11-04; note that our methods do not attempt to use common properties of homologous flares in this way.
\nocite{2005JGRA..110.9S00G}
\cite{2013JSWSC...3A..31C} had previously combined the Kiplinger and Garcia X30.6 classification with peak fluxes inferred from sudden ionospheric disturbances for this event \citep[e.g.,][]{2005JGRA..110.6306T} to obtain a GOES classification of X35$\pm$5.  
The Kretzschmar-Schrijver relationship between flare total solar irradiance and GOES class yields an $\sim$X35 classification for this flare based on its bolometric energy of $4.3 \times 10^{32}$ erg \citep{2012ApJ...759...71E,2022LRSP...19....2C}.    
From the linear analysis of the events in Table~\ref{tab:test}, we would expect the inferred values to be about 10\% lower on average than the unknown true values and our peak times to be 1-2 minutes earlier.
The ~10\% underestimates of peak GOES fluxes given by the linear method make it a conservative approach for the reconstruction of the high-flux end of the ODF described in Section~\ref{sec:ofd} below.

\begin{table}[h]
\caption{Saturated Events}
\label{tab:meth}
\begin{tabular}{l l c c r c c}
\hline
Event & Satellite & Class$^*$ & Maximum & Duration & Linear & Functional \\
           & &  &10$^{-4}~$Wm$^{-2}$ & mm:ss & Class & Class \\
\hline

SOL1978-07-11 &GOES-2 & X15  &  1.57 & 11:05 & X31.2  & X33.1 \\
SOL1984-04-24 &GOES-5 & X13 &  1.30  &  3:42 & X13.9  & X13.7 \\
SOL1989-03-06 &GOES-6& X15 &  1.18  &  7:06 & X13.5  & X13.2 \\
SOL1989-08-16  &GOES-6& X20 &  1.16  & 32:23 & X20.5 &  X18.8 \\
SOL1989-10-19  &GOES-6& X13 &  1.16  & 19:32 & X17.0  & X15.4\\
SOL1991-06-01 &GOES-6 & X12 &  1.16  & 26:38 & X20.1  & X19.6\\
SOL1991-06-04  &GOES-6& X12 &  1.16  & 20:31 & X23.1  & X21.7\\
SOL1991-06-06 &GOES-6 & X12 &  1.16  & 16:19 & X23.1  & X19.2\\
SOL1991-06-11  &GOES-6& X12 &   1.16  &   3:04 & X12.0  & X11.9  \\
SOL1991-06-15  &GOES-6& X12 &  1.16  & 11:38 & X25.3  & X23.3\\
SOL2001-04-02  &GOES-10& X20 &  1.84  & 4:12 & X21.4  & X20.1\\
SOL2003-11-04  &GOES-10& X28 &   1.84  & 13:36 & X30.5  & X30.0\\
\hline
\end{tabular}
$^*$\url{https://www.ngdc.noaa.gov/stp/space-weather/solar-data/solar-features/solar-flares/x-rays/goes/xrs/}
\end{table}

\subsubsection{Functional fits}\label{sec:functional}

\begin{figure}[htbp]
\centering
   \includegraphics[width=0.9\textwidth]{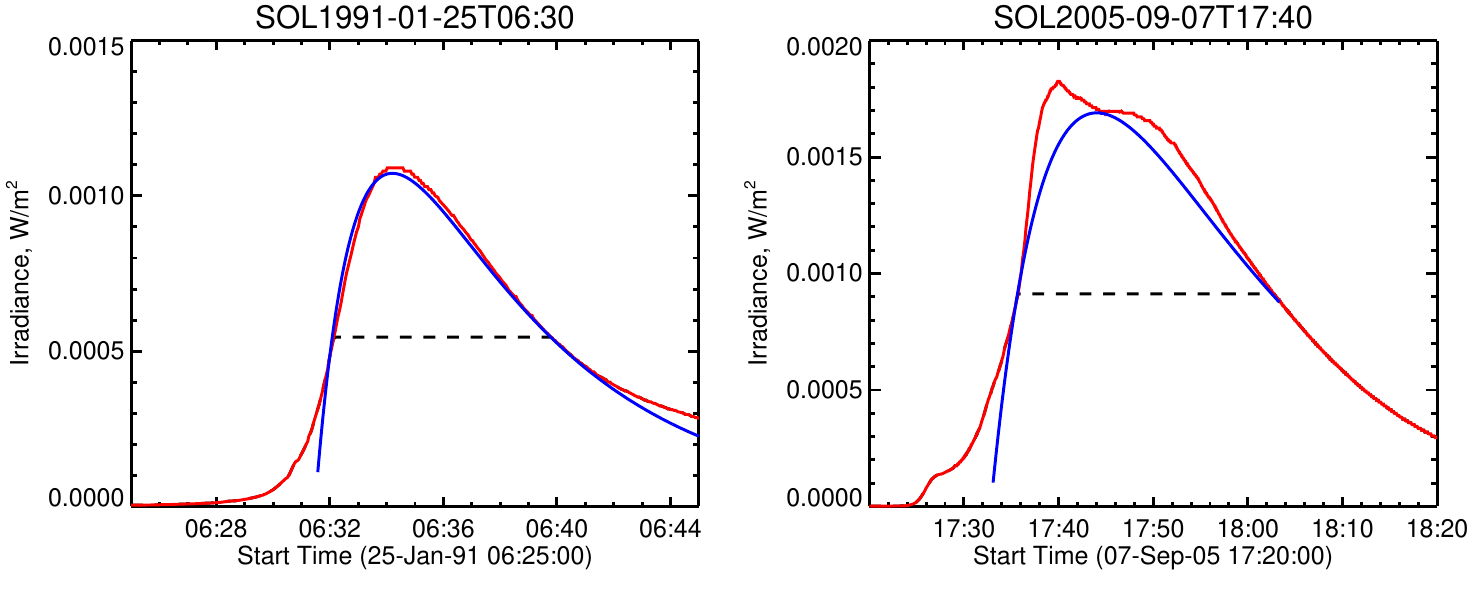}
          \caption{\textit{Time series for 1-8~\AA\ for the same two reference events shown in Figure~\ref{fig:f1},
          but now with the functional fits (Equation~\ref{eq:function}) superposed (blue) on the artificially truncated GOES data (red).
          The right panel (SOL2005-09-07) shows a case with an error in peak flux of about 8\%, and a 3-min miss in time of peak.
         } 
      }
\label{fig:functional_fit_plots}
\end{figure}

Because of the simplicity of the linear fits, we repeated the curve-fitting exercise across the data gaps in the saturated events, now using a functional form more representative of flare soft X-ray time profiles. 
Thus we fitted a smooth-topped characteristic function described by
\begin{equation}\label{eq:function}
F(t) = A\,(1-e^{-\Delta t/\tau_1})\,e^{-\Delta t/\tau_2}\  \ ,
\end{equation}
with $\Delta t$  measured relative to an adjustable reference time $\tau_0$, nominally the peak time as determined by a bisector method. 
This is basically a four-parameter fitting procedure, applied manually with reference to subjective good\-ness-of-fit to the function.  
Because of the rounded peak, we would expect to find slightly lower peak fluxes this way, and this is indeed the case, with the median estimated peaks ranging from 70\% to  103\% of the actual peaks, with a median value of 89\%. 
Figure~\ref{fig:functional_fit_plots} shows the timeseries for the two events shown in Figure~\ref{fig:f1}.

\subsubsection{Composite estimates and uncertainties}\label{sec:uncertainties}

The test events (Table~\ref{tab:test}), as fitted with the two methods described above, give a direct estimate of uncertainties, with the estimated peak GOES fluxes typically  
$\approx$~10\% below the actual values, with a full range for both methods from $\geq$30\% to $\geq$10\%, comparable to the GOES calibration errors which (as discussed in Section~\ref{sec:recal}) could be as large as 20\% from satellite to satellite.  
Table~\ref{tab:meth} compares the methods and gives an (rms) difference for our extrapolations of the saturated events of a few percent.
We emphasize that these two methods represent no more than empirical approximations, done independently by authors Cliver (linear) and Hudson (functional).
We note that the uncertainty associated with the extrapolation technique is probably smaller than that of the GOES calibration itself, which (as discussed in Section~\ref{sec:recal} above) could be as large as 20\% from satellite to satellite.
The two methods generally agree; we have taken their average as our estimate and give statistics in Table~\ref{tab:stats}.

\begin{table}[h]
\caption{Comparison of the fitting methods by ratio}
\label{tab:stats}
\begin{tabular}{l r r r r r}
\hline
Event Set &  Mean  & Median & Standard & Range \\
 &    &  & Deviation &  \\
\hline
Linear/Functional  &    1.06  &  1.02 & 0.09  &  0.93 - 1.23\\
Average/Actual  &    0.90  &  0.91 & 0.09  &  0.74 - 1.02\\
\hline
\end{tabular}
\end{table}

\section{Occurrence Frequency Distributions}\label{sec:ofd}

\subsection{Database}

The study of occurrence distribution functions (ODF) for coronal flare signatures may have originated with \cite{1956PASJ....8..173A}, who studied the distribution of peak fluxes of microwave bursts.
These showed a clear power-law distribution $N(E) \propto E^{-\gamma}$, with $\gamma = 1.8$. 
Such a power law has subsequently been found for many observables in solar (and stellar) flares; these distributions are often taken to represent distributions of total flare energy \citep[e.g.,][]{2012ApJ...759...71E}, 
but are always only proxies because of data incompleteness, and
as such they may have significant biases across a distribution of magnitudes. 
At values of $\gamma$ below 2.0, a power law might imply a divergence in the total flare energy summed over all events \citep{1991SoPh..133..357H}. 
Accordingly either the interpretation of the observable as a proxy for total flare energy must be flawed, or else the power-law extension to extremely powerful events must truncate at some energy level. 
In this context the GOES soft X-ray data have particular importance. 
Again though, they represent only a small fraction of the total event energy, of order a few percent.
In principle the recent observations of total solar irradiance (TSI) allow for a better characterization of flare energy via summed-epoch analysis \citep{2011A&A...530A..84K} even though only a small number of individual events can be detected directly in this way \citep{2006JGRA..11110S14W,2014ApJ...787...32M}.

The GOES ODF has been studied numerous times previously \citep[e.g.,][]{1997ApJ...474..511F,2002A&A...382.1070V,2012ApJ...754..112A,2012ApJS..202...11R,2022Physi...5...11S,2023AdSpR..71.2048P}, with varying results.
Many other studies of flare ODFs in different observables exist, often described as proxies for total flare energy, but most of them describe energetically insignificant quantities and are difficult to intercompare for this reason.
Our approach improves on some of the previous ones, as described below, and also of course is more complete because it extends to the time of writing. 

With the revised values for the saturated events, we now have a complete sample of the peak soft X-ray fluxes between September 1975 and June 2022, a total of 48,131 at listed classes C and above\footnote{\url{https://umbra.nascom.nasa.gov/goes/fits/}}.
As noted above, these original metadata include errors and ambiguities. 
This would be expected in view of the operational nature of the data, which involved substantial human intervention in styles that have not been documented uniformly over the decades. 
In addition, timewise overlapping flares frequently occur at high activity levels, and this increases the uncertainty of the original classes interpreted as peak flux levels in those cases.
To systematize the data, we have used the existing SolarSoft access 
framework as embodied in {\sc ogoes} to obtain full data files, as maintained at time of writing by NOAA and in the calibration on the original scale prior to GOES-13.
Crucially the original metadata list the flare magnitudes, without background correction, that were obtained through 
1-minute averages.
In the present work we use the SolarSoft guidelines to establish a uniform re-reduction of the data as described before, leading to our independent derivation of the ODF for flare excess fluxes.
The ODF thus derived has the scaling factor applied to the earlier data so that the entire database matches the calibration of data obtained from the GOES-16 through GOES-18 spacercraft at the time of writing.

From the event list we have extracted the time-series data at full resolution (3, 2, or 1~s depending upon epoch).
The SolarSoft \citep{1998SoPh..182..497F} access tool for GOES data, by default, provides data from the primary GOES satellite in the common case of redundant coverage, and we use only these data.
For each event we use an algorithm to estimate the peak measured flux and a corresponding preflare background value, almost always at the time of the Channel-A (0.5-4~\AA) minimum within the 30-min interval prior to the Channel-B (1-8~\AA) peak.
The resulting ODF therefore contains flare excess fluxes on the assumption that this chosen background  level represents other (non-flaring) X-ray sources \citep[see][for a discussion]{1990ApJ...356..733B}. 
Note that such event-by-event background estimates have also been carried out by \cite{2012ApJ...754..112A} in an earlier study of the first 37 years of these data; other independent studies have made no background adjustments \citep[\textit{e.g.,}][]{2002A&A...382.1070V} or instead have used global estimates of a background level, rather than flare-specific ones \citep[e.g.,][]{2012ApJS..202...11R}. 
\cite{2012PhDT........57H} describes an algorithmic background definition for EVE \citep{2012SoPh..275..115W}, also a Sun-as-a-star instrument.
Data reduction without event-by-event background correction generally introduces systematic biases at lower peak flux levels and are therefore not suitable for the ODF calculation. 
Even with an event-by-event background adjustment, there still remains uncertainty systematically larger for the lower peak flux levels. 
The effects of this on the ODF generation cannot be known \textit{a priori} and we assume here that this factor is unbiased if not negligible, our main interest being in the X-class events.

\subsection{Fitting the powerlaw region}\label{sec:powerlaw}

\begin{figure}[htbp]
\centering
   \includegraphics[width=0.55\textwidth]{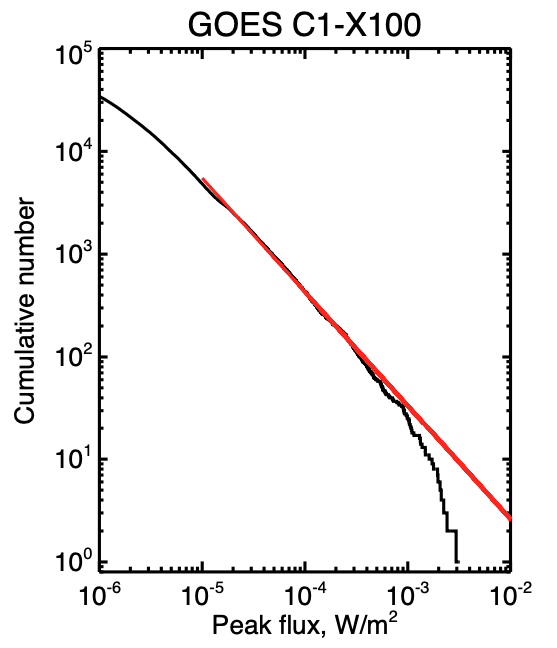}
  \includegraphics[width=0.39\textwidth]{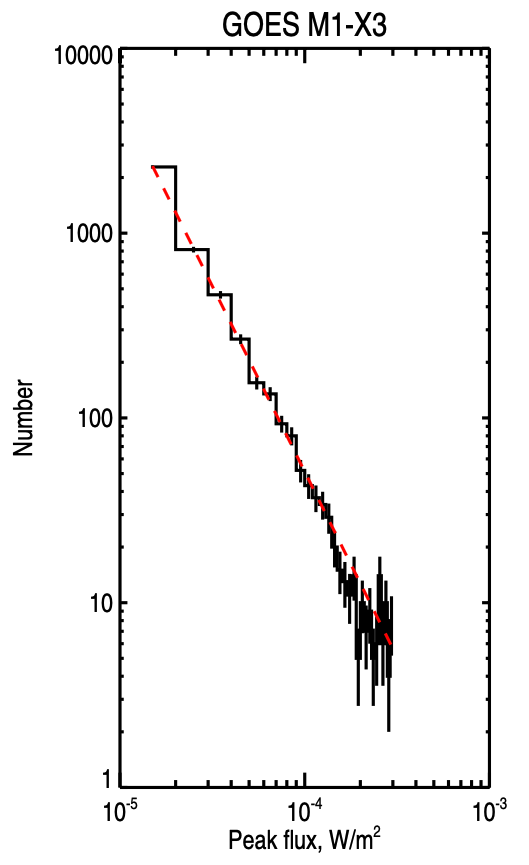}
          \caption{\textit{Left, downward-cumulative distribution of 47,173 GOES soft X-ray flares (C1 and above) over 1975--2022, in bins of $10^{-7}$~W/m$^2$ (B1) width.
          Right, differential distribution fitted over the range M1--X3 only, with a red line showing a least-squares fit to the M1-X3 range.
          The (extrapolated) red line in the left panel comes from the same fit.
         } 
      }
\label{fig:cumulative}
\end{figure}

Figure~\ref{fig:cumulative} shows the downward-cumulative distribution function (DCDF) for the entire sample. 
Using the maximum-likelihood method over the range M1 through X3, we find that peak fluxes $F$ follow a power-law occurrence distribution of the form $F \propto E^{-\gamma}$, with index $\gamma$ = 1.973$\pm0.014$ (see below), with interesting implications. 
First, this confirms the result of \cite{2012ApJ...754..112A} that flare soft X-ray fluxes appear to have a steeper distribution than some other parameters that might serve as proxies for the total event energy, statistically consistent with a power-law slope $\gamma$ = 2.00.
This is inevitably an uncertain point, since the peak flux of an event (and any other proxy, by definition) does not actually measure its energy directly.
Second, this is the marginal value for a limited total energy over the ensemble \citep{1991SoPh..133..357H}.
Third, as discussed below, the GOES data clearly require a truncation of the powerlaw at about X10.

The cumulative distribution contains 6,344 events above a background-sub\-tracted peak flux 10$^{-5}$~W/m$^2$ (M1) and has a power-law appearance at the lowest range. 
We fit the range up to the equivalent X3 flux as shown in Table~\ref{tab:fits}, using the maximum-likelihood method \citep[e.g.,][]{1970ApJ...162..405C,2016SoPh..291.1561D}.
The binned distribution (Figure~\ref{fig:cumulative}, right panel) shows a consistent fit, slightly steeper as expected, and we adopt its uncertainty value for extrapolations.
The slope is constent with 2.0 but without background subtraction this would not be the case; the slope of the power-law range of the ODF is significantly steeper in this case.

\begin{table}
\caption{Power-law $\alpha$ (differential) over equivalent M1-X3 range}\label{tab:fits}
\begin{tabular}{l l l}     
\hline
Maximum Likelihood & $1.973 \pm 0.014$ & Backgrounds subtracted \\
Maximum Likelihood & $2.148 \pm 0.015$ & Backgrounds not subtracted \\
Binned			& 2.008 $\pm\  0.048$ & Backgrounds subtracted \\
Binned			& 2.172 $\pm\  0.043$ & Backgrounds not subtracted \\
\hline
\end{tabular}
\end{table}

The ODF shown in Figure~\ref{fig:cumulative} (left panel) shows a deficit of major events, and because the GOES database has essentially complete sampling over its 46 years to time of writing, we can now discuss its implications.
We have no theoretical guidance for any particular form for the rollover and do not attempt here to characterize it with an empirical formula, such as a power law with an exponential cutoff or the  alternative of a broken power law \citep[e.g.,][]{1993ApJ...413..281B}, which would have distinctly different predictions for longer time scales.
Instead we just characterize the significance of the rollover by extrapolating the observed power law from the weaker events (below X3), with $\alpha = 1.973$, as described in Table~\ref{tab:numbers} (left columns).
The distributions (observed and predicted, based on the results in Table~\ref{tab:fits}) clearly disagree strongly, with the simple power law significantly overestimating the event numbers in the observable range.
The inclusion of the historical Carrington event itself \citep[e.g.,][]{2013JSWSC...3A..31C,2021ARA&A..59..445H} could not alter this conclusion significantly based on the equivalent GOES class inferred from proxy data in these papers.  \cite{2010...NAM...C} do suggest a value of about X42 (old scale, without the factor 1.43) based on this flare's geomagnetic effect \citep{1861RSPT..151..423S}.
Recently \cite{2023ApJ...954L...3H} also suggested X56 (old scale) based on re-analysis of the original optical sighting.
We note the similarity of the radio-burst ODF discussed by \cite{2002ApJ...570..423N}, who also found a paucity of great events (noting as well the possibility of a contribution by instrumental saturation).

\begin{table}
\caption{Event numbers normalized at X1}\label{tab:numbers}
\begin{tabular}{ccccc}     
\hline
Class & Observation & Expected number & \textit{Predicted number} & \textit{\cite{2018_201837G}}\\
                &                    & (M1-X3 power law) & \textit{(this work)} & (\textit{power law $\alpha = 2.31$})\\
\hline
X1             &  609          &609   &\textit{609} &\textit{609} \\
X3            &  174           &208  & \textit{169}& \textit{47} \\
X10             &  37          & 64 & \textit{35} & \textit{2}\\
X30             &  4            & 22   & \textit{5} & \textit{0}\\
X100           &  0            & 14    & \textit{0} & \textit{0}\\
X300         &  0              &       & \textit{0} & \textit{0}\\
\hline
\end{tabular}
\end{table}

\section{Fitting the CCDF: the ``superflares''}\label{sec:extrapolation}

\begin{figure}[h]
\centering
   \includegraphics[width=\textwidth]{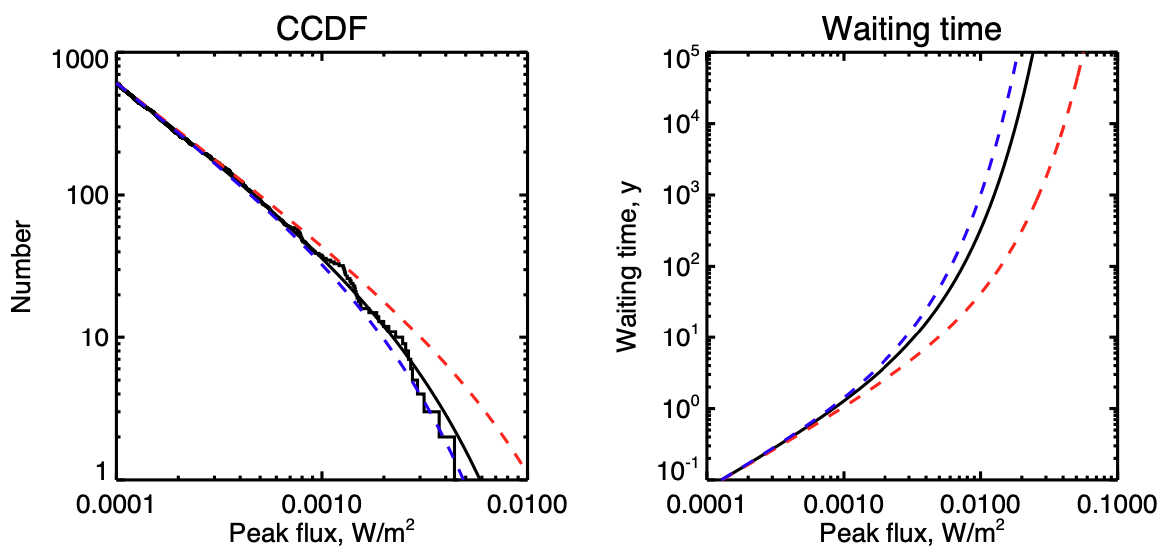}
          \caption{\textit{Left, downward-cumulative distribution of 609 GOES soft X-ray flares (new X1 level and above) over 1975--2022.
          Right, the same information inverted to waiting time for a flare at a given GOES class.
          See text for details.} 
      }
\label{fig:prediction_final}
\end{figure}

The complete sample now lets us fit empirical functions to the GOES complementary cumulative occurrence distribution function (CCDF). 
By doing this we can have a mathematical basis for extending to still greater events than have been observed, such as  the SXR flares inferred for the cosmogenic (radioisotope-based) solar energetic proton (SEP) events \citep{2012Natur.486..240M,2013A&A...552L...3U}, with others  now noted discovered or identified across the time scale of the Holocene \citep{2022LRSP...19....2C,2023LRSP...20....2U}. 
Superflares observed on other stars are arbitrarily defined as those with bolometric energy $> 10^{33}$~erg \citep{2000ApJ...529.1026S}. 
In comparison, the largest flares directly observed on the Sun have radiative energies of at most $\approx 4 \times 10^{32}$~ erg \citep[e.g.,][]{2012ApJ...759...71E}.
The departure of the observed occurrence distribution (the ODF) from a simple power law complicates this extension, which must now have additional parameters to describe the observed roll-over, just where small numbers lead to uncertain statistics.
We attack this problem by fitting the empirical CCDF to one of the mathematical forms used by \cite{2022Physi...5...11S} who tested four different fits: a simple power law (one free parameter), tapered power law (two), gamma function (two), and Weibull (two). 
As noted by Sakurai, the Weibull distribution does not fit the roll-over (required by the lack of extreme events) as well as the other two-parameter fits, and for simplicity we use only the tapered power law here:
\begin{equation}\label{eq:ccdfpwrexp}
CCDF = A (S/S_0)^{-\alpha-1} e^{-\beta(S-S_0)/S_0}\  ,
\end{equation}
representing the CCDF of $S$ (here in W/m$^2$) above a threshold peak flux S$_0 = 10^{-4}$~W/m$^2$ (the X1 level). 
Note that this (or any other) functional form for the CCDF is purely a mathematical convenience since we have no particular physical motivation for any of the options. 
We wish to estimate the uncertainties the CCDF fit function over the observed range.
Possible goodness-of-fit approaches include $\chi^2$ and Kolmogorov-Smirnov tests \citep{2022Physi...5...11S} for the entire distribution, but we choose instead simply to freeze the $\alpha$ parameter at Sakurai's value, which agrees with our maximum-likelihood fit to the M3--X10 range as described in Section 3.2. 
This means that our estimated uncertainties depend only on the statistics of the roll-over region itself, namely the X10-class events.
We estimate uncertainties on the $\beta$ parameter by assuming a Poisson distribution for events at X10 class or above, for which we find a total of 37$\pm$6 events, leading to parameters $[\alpha, \beta] = [0.973, 0.0336^{+0.011,-0.021}]$. 
Figure~\ref{fig:prediction_final} shows the results in terms of the CCDF and the waiting time vs. GOES peak flux. This can be crudely transformed into total event energy $E$ by using Eq.~5 of \cite{2022LRSP...19....2C},  $E = 3.3 \times 10^{31}(S/S_{X1})^{0.72}$~erg as scaled to the GOES peak flux $S$ in units of $10^{-4}$~W/m$^2$. 
The upper limit for the total energy of a solar flare on the Holocene time scale of about 10$^4$ years thus becomes $\approx$10$^{33}$ erg. 

We note that CCDF extrapolations by \cite{2018_201837G} and by \cite{2022Physi...5...11S}, based on samples of unscreened GOES flares with no correction for the saturated (clipped) events, obtained similar results. 
Using a Weibull function fit, \cite{2018_201837G} obtained 100-yr and 1000-yr estimates of X63 and X144, respectively (with the 1.43 correction applied) while \cite{2022Physi...5...11S} estimated that a once-per-1000-yr flare would have a X100 GOES class. 
Table~\ref{tab:numbers} also shows an alternative power-law extrapolation to extreme values (Gopalswamy 2018, Fig.~5), along with our tapered power law, in the right two columns.
These values compare with the our nominal 100-yr (X75) and 1000-yr (X110) estimates in Table~\ref{tab:return}.  
All of these estimates fall short of the (current scaling) $\approx$X$400\pm200$ ($\approx 2 \times 10^{33}$~erg) estimate obtained by \citep[][cf. Cliver \textit{et al.} 2022]{2020ApJ...903...41C} for the cosmogenic-nuclide-based solar energetic particle (SEP) event of 774~AD with an inferred $>$200 MeV fluence 80$\times$ larger than the largest SEP event of the modern era, that associated with the flare SOL1956-02-23 \citep{2021A&A...647A.132K,2023JGRA..12831186K}.  

From the CCDF in Figure~\ref{fig:prediction_final} (left), we can determine the GOES event occurrence on 10-, 100-, and 10,000-yr time scales from the corresponding waiting-time plot on the right of the figure, which we interpet in Table~\ref{tab:return}.

\begin{table}[h]
\caption{Probable event magnitudes for long time scales}\label{tab:return}
\begin{tabular}{rcc}     
\hline
           &  GOES class & Bolometric Energy$^*$ \\
           &                       & ($10^{32}$~erg) \\
 \hline
100-yr            &  X60-\textbf{X75}-X115 &4.9-\textbf{5.7}-7.8   \\
1000-yr            &  X100-\textbf{X110}-X260 &7.0-\textbf{7.5}-14.0   \\
10$^4$-yr            &  X140-\textbf{X180}-X400 &7.8-\textbf{10.7}-19.1   \\
\hline
\end{tabular}

$^*$Based on the Kretzschmar-Schrijver relationship between\\ GOES class and bolometric
energy, Eq.~5 in \cite{2022LRSP...19....2C},\\ with the adjustment for the new GOES as described\\ in Section~8.2.1 therein.

\end{table}

\section{Conclusions}\label{sec:conclusions}

In this analysis of GOES SXR events from 1975-2022, we have estimated the peak fluxes for 12 1-8~\AA\ events of class $\geq$X10 that saturated (or more accurately were clipped) prior to reaching their maximum.  
We did so by developing a technique of extrapolating the clipped time profile across the saturation gap by two independent methods that we verified on a set of 10 $\geq$X10 non-saturated events.
In addition, we applied a significant calibration correction (43\% increase in 1-8~\AA\ peak fluxes) to the GOES-1 through GOES-15 data (1975-2016) and also subtracted an event-by-event background for all flares with GOES class $\geq$C1.
For events below the saturation range, we fitted the ODF with a power-law with index near 2.0, consistent with the results of \cite{2012ApJ...754..112A}
We have now included the saturated events by a uniform treatment of their extrapolated time profiles, as verified on a set of non-saturating events.
The observed ODF departs from its power-law  character at the high end; extrapolation to higher GOES fluxes beyond the $\approx$X10-level over-predicts the observed occurrence frequency of such events.
 
Extrapolation of a tapered-power-law fit to the GOES 1-8~\AA\ CCDF yields a nominal (revised SWPC scaling) 100-yr event of X75, comparable to a recent $\approx$X80 assessment of the Carrington event \citep{2023ApJ...954L...3H}, and a 1000-yr event of X110, bracketed by recent (new-scale) estimates of X100 \citep{2022Physi...5...11S} and X144 \citep{2018_201837G} based on GOES event samples without event-screening, back-ground subtraction, or saturation correction.  
The agreement of these various studies suggests a certain robustness for such estimates. 
Our nominal 1000-yr estimate of X180 (X140-X400), however, falls outside the estimate of X400$\pm$200 peak GOES flux \citep{2020ApJ...903...41C,2022LRSP...19....2C} for the flare associated with the huge cosmogenic-nuclide-based SEP event of 774~CE \citep{2021A&A...647A.132K,2023JGRA..12831186K}, although the uncertainties overlap.
We note that such comparisons involve a complicated chain of inference (solar event~$\rightarrow$~soft X-rays~$\rightarrow$~CME~$\rightarrow$~SEPs~$\rightarrow$~Earth impact~$\rightarrow$~radioisotope record), each step of which involves assumptions.

\begin{acks}
The authors have benefited from participation at several ISSI Team meetings.
This comprehensive study would not have been possible without the careful work of our late friend and colleague, Richard Schwartz.
\end{acks}

\bigskip
\noindent{\bf Appendix: The X10-class GOES events, revised}\label{sec:appendix}

\bigskip
\noindent Table~\ref{tab:original} provides new GOES peak-flux estimates for 
23~X10-class events (original listing) in the database, and Table~\ref{tab:augmentation} extends it to a total of 38 according to the correction factor of 1.43 applied to the 1-8~\AA\ magnitudes.
The sorting here is by our new estimates; note that only event No.~19 (X9.3) did not have an original GOES class at $\geq$X10 (though GOES-15 and GOES-16 differed on this point).
The new magnitudes adjust peak flux estimates from GOES-12 through GOES-15 and earlier to match the calibrations of the GOES-R series.

The listed peak fluxes do not exactly correspond to the older class ratings, based on 1-minute sampling, but correspond almost exactly at two significant figures: our top entry of 45.9$ \times 10^{-3}$~W/m$^2$ can be taken as X46 (or X32 on the old scale).
We note that the background corrections, though important in characterizing the ODF, normally do not matter for the greatest events.
These tables are in order of our new estimation of GOES peak flux levels, incorporating the event-by-event background correction, and using the full time resolution of the sampling rather than 1-min integrations.
According to our validation procedure for the corrections to the saturated events (Section~\ref{sec:uncertainties}) we believe that the individual events have peak fluxes estimated with an rms precision of about 3\%, which is small compared with the nominal XRS radiometric error (see Figure~\ref{fig:rats_plot}), which we have estimated at 20\%.
SOL1978-07-11T10:58 has a compromised calibration since GOES-2 did not have sufficient overlapping in-orbit data from other GOES instruments.
Most likely the GOES peak flux for SOL2003-11-04 is a lower limit, since the flare was a limb event and somewhat occulted, so this event is probably the greatest GOES event in our database.

\begin{table}[h]
\caption{The greatest GOES events (original)}\label{tab:original}
\begin{tabular}{ r r r l c l }
\hline\noalign{\smallskip}
Rank & IAU Identifier & Old class & Satellite &  Saturated? & Peak Flux \\
&&&&Y/N&10$^{-3}$~W/m$^2$ \\
\hline
           1&SOL1978-07-11T10:58&X15.0&GOES2&Y&X45.9\\
           2&SOL2003-11-04T19:50&X28.0&GOES10&Y&X43.2\\
           3&SOL1991-06-15T08:31&X12.0&GOES6&Y&X34.7\\
           4&SOL1991-06-04T03:47&X12.0&GOES6&Y&X32.0\\
           5&SOL1991-06-06T01:12&X12.0&GOES6&Y&X30.2\\
           6&SOL2001-04-02T21:51&X20.0&GOES8&Y&X29.6\\
           7&SOL1991-06-01T15:29&X12.0&GOES6&Y&X28.3\\
           8&SOL1989-08-16T01:17&X20.0&GOES6&Y&X28.0\\
           9&SOL2003-10-28T11:10&X17.2&GOES10&N&X25.7\\
          10&SOL2005-09-07T17:40&X17.0&GOES12&N&X24.6\\
          11&SOL1989-10-19T12:55&X13.0&GOES6&Y&X23.1\\
          12&SOL2001-04-15T13:50&X14.4&GOES8&N&X21.1\\
          13&SOL1984-04-24T24:01&X13.0&GOES5&Y&X19.7\\
          14&SOL1989-03-06T14:10&X15.0&GOES6&Y&X19.3\\
          15&SOL1982-12-15T01:59&X12.9&GOES2&N&X19.0\\
          16&SOL1991-06-11T02:29&X12.0&GOES6&Y&X17.0\\
          17&SOL1991-01-25T06:30&X10.0&GOES6&N&X15.7\\
          18&SOL2003-10-29T20:49&X10.0&GOES10&N&X15.5\\
          19&SOL1991-06-09T01:40&X10.0&GOES6&N&X15.1\\
          20&SOL1984-05-20T22:41&X10.1&GOES5&N&X14.8\\
          21&SOL2017-09-06T12:02&X9.3&GOES16&N&X14.8\\
          22&SOL1982-06-06T16:54&X12.0&GOES2&N&X14.7\\
          23&SOL1982-12-17T18:58&X10.1&GOES2&N&X14.7\\
          \end{tabular}
          
$^*$The GOES calibration for this early event is relatively weak; see Figure~\ref{fig:rats_plot}          
\end{table}

\begin{table}[h]
\caption{The greatest GOES events (augmentation)}
\label{tab:augmentation}
\begin{tabular}{ r r r l c l }
\hline\noalign{\smallskip}
Rank & IAU Identifier & Old class & Satellite & Peak flux \\
&&&&10$^{-3}$~W/m$^2$ \\
\hline
          24&SOL1989-09-29T11:33&X9.80&GOES6&X14.2\\
          25&SOL1990-05-24T20:49&X9.30&GOES6&X14.1\\
          26&SOL1991-03-22T22:45&X9.40&GOES6&X13.8\\
          27&SOL2003-11-02T17:25&X8.30&GOES10&X13.3\\
          28&SOL1997-11-06T11:55&X9.40&GOES8&X13.1\\
          29&SOL2006-12-05T10:35&X9.00&GOES12&X13.1\\
          30&SOL2017-09-10T16:06&X8.2&GOES16&X13.0\\
          31&SOL1992-11-02T03:08&X9.00&GOES6&X13.0\\
          32&SOL1980-11-06T03:48&X9.00&GOES2&X12.8\\
          33&SOL1982-06-03T11:48&X8.00&GOES2&X11.8\\
          34&SOL1989-10-24T18:31&X5.70&GOES6&X11.5\\
          35&SOL2011-08-09T08:05&X6.90&GOES15&X10.7\\
          36&SOL1991-03-04T14:03&X7.10&GOES6&X10.4\\
          37&SOL2005-01-20T07:01&X7.10&GOES12&X10.2\\
          38&SOL1982-07-12T09:18&X7.10&GOES2&X10.1\\
\end{tabular}
\end{table}

\clearpage
\bibliographystyle{spr-mp-sola}
\bibliography{greatest.bib}  


\end{article} 
\end{document}